\documentclass[letter]{IEEEtran}

\usepackage{amsmath,amssymb,epsfig,cite,footnote,authblk}
\usepackage{color,hyperref}

\hypersetup{
    colorlinks,%
    citecolor=blue,%
    filecolor=blue,%
    linkcolor=blue,%
    urlcolor=blue
}
\definecolor{bluecolor}{rgb}{0,0.,1.}

\definecolor{redcolor}{rgb}{.7,0.,0.}

\newcommand{\pr}[1]{\left( #1\right)}
\newcommand{\prr}[1]{\left[ #1 \right]}
\newcommand{\es}[1]{\begin{equation}\begin{split}#1\end{split}\end{equation}}

\newcommand{\R}{\mathbb{R}}

\newcommand{\V}{\mathcal{V}}

\newcommand{\dd}{\textrm{d}}

\newcommand{\squeezeup}{\vspace{-1.5mm}}
\begin{document}

\title{\huge{Low Power Wide Area Network Analysis: Can LoRa Scale?}}
\author{Orestis Georgiou}
\author{Usman Raza}
\affil{\small{Toshiba Telecommunications Research Laboratory, 32 Queens Square, Bristol, BS1 4ND, UK} \thanks{orestis.georgiou@toshiba-trel.com}}
\maketitle


\begin{abstract}
Low Power Wide Area (LPWA) networks are making spectacular progress from design, standardisation, to commercialisation. 
At this time of fast-paced adoption, it is of utmost importance to analyse how well these technologies will scale as the number of devices connected to the Internet of Things (IoT) inevitably grows. 
In this letter, we provide a stochastic geometry framework for modelling the performance of a single gateway LoRa network, a leading LPWA technology. 
Our analysis formulates unique peculiarities of LoRa, including its chirp spread-spectrum modulation technique, regulatory limitations on radio duty cycle, and use of ALOHA protocol on top, all of which are not as common in today's commercial cellular networks. We show that the coverage probability drops exponentially as the number of end-devices grows due to interfering signals using the same spreading sequence. 
We conclude that this fundamental limiting factor is perhaps more significant towards LoRa scalability than for instance spectrum restrictions. 
Our derivations for co-spreading factor interference found in LoRa networks enables rigorous scalability analysis of such networks.
\end{abstract}

\squeezeup\section{Introduction \label{sec:intro}}

The last years have seen  much interest in Low Power Wide Area (LPWA) technologies, which are gaining unprecedented momentum and commercial interest towards the realisation of the Internet of Things (IoT).
There are many candidates that have taken the research community by surprise, actively pursuing standardisation, adoption, and commercial deployments in parallel.
Most LPWA networks operate in the unlicensed ISM bands at 169, 433, 868/915 MHz, and 2.4 GHz depending on the region of operation.
Some of the most pronounced LPWA candidates are SigFox, LoRa, Weightless, and Ingenu \cite{raza2016low}.

In this paper, we focus on LoRa (Long Range), one of the most promising wide-area IoT technologies proposed by Semtech and further promoted by the LoRa Alliance \cite{sornin2015lorawan}. 
At the heart of LoRa's success is its adaptive data rate chirp modulation technology allowing for flexible long-range communication with low power consumption and low cost design. 
Essentially, this is achieved via spread spectrum multiple access techniques accommodating multiple users in one channel. 
LoRa Alliance has defined the higher layers and network architecture on top the LoRa physical layers and termed them LoRaWAN. 
Together, these features make LoRa attractive to developers who can build complete system solutions on top of it for both geographical and residential/industrial types of IoT networks, thus fast-tracking its market adoption.

Despite this success, LoRa has not yet attracted similar levels of attention from the academic and research community with only very few peer-reviewed studies published to date \cite{raza2016low,vangelista2015long,goursaud2015dedicated,
petajajarvi2015coverage,augustin2016study,
margelis2015low}.
In fact, most of these studies are either review articles \cite{raza2016low,vangelista2015long,goursaud2015dedicated}, or measurement reports \cite{petajajarvi2015coverage,augustin2016study}.
Significantly, it has consistently been assumed in peer-reviewed and industry published white papers that the different spreading sequences employed by LoRa are orthogonal and provide interference immunity at the receiver end.
{While this technology does indeed create an extra set of ``virtual" channels thus increasing the capacity of the gateway, transmissions of similar spread are susceptible to a new type of interference, unique to LoRa networks, which we term \textit{co-spreading factor interference}.
Since LoRaWAN's MAC protocol is essentially an ALOHA variant with no collision avoidance provisions \cite{vangelista2015long}, in very dense deployment scenarios, LoRa networks will inevitably become interference-limited, rather than noise-limited, thus necessitating for new interference-related performance metrics which capture the interference peculiarities of LoRa networks.}

It is the purpose of this letter to apply state-of-the-art mathematical tools to model the uplink coverage of single gateway LoRa networks and further understand its unique PHY and MAC features.
Namely, we leverage tools from Stochastic Geometry \cite{haenggi2012stochastic} to address { two independent link-outage conditions, one concerned with SNR}, and the other with co-spreading factor interference.
{We show that LoRa is susceptible to the latter, and that end-device coverage probability decays exponentially with increasing number of end-devices}, despite  the `cushioning' provided by the low duty cycling and chirp orthogonality.

\squeezeup\section{LoRa Basics \label{sec:Lora}}

The LoRa network operates in the sub-GHz band. 
Maximum transmit powers are defined as 14 and 21.7 dBm in Europe and the US, respectively.
The LoRa system consists of end-devices\footnote{There are 3 classes of end-devices: Class A (for All), B (for Beacon) and C (for Continuously listening), each associated to a different operating mode.}, gateways, and the NetServer forming a star of stars topology with the NetServer at the root, the gateways at level one, and end-devices as the leaves \cite{sornin2015lorawan}.
The PHY and MAC layers of LoRa were recently described in \cite{goursaud2015dedicated}.
At the heart of LoRa is a proprietary chirp spread spectrum (CSS) modulation scheme.
For binary chirp modulation, the data passes through a chirp modulator that maps each bit chunk to 1 of 2 waveforms:
\es{
s(t) \!=\! \sqrt{\frac{2E_s}{T_s}} \cos\big[ 2 \pi f_c t \!\pm\! \pi \big( u\big(\frac{t}{T_s}\big) \!-\! w \big( \frac{t}{T_s}\big) ^2 \big) \big] 
,}
where $E_s$ is the energy of $s(t)$ in the symbol duration $T_s$, $f_c$ is the carrier frequency, and the parameters $u$ and $w$ are the peak-to-peak frequency deviation, and the sweep width, respectively, both normalised by the symbol rate.
Notably, LoRa supports variable/adaptive data rate, thus enabling the trade-off between throughput for coverage range, or robustness, or energy consumption, while keeping a constant bandwidth.
This is managed by the NetServer by regulating the bandwidth $\text{BW}$ and the spreading factor $\text{SF}\!\in\!\{7,8,\ldots,12\}$ in Europe which determines the length of the chirp symbol $T_s\!=\!2^{\text{SF}}/\text{\text{BW}}$.
As such, the symbol duration and hence the time-on-air of a transmission increases exponentially with $\text{SF}$ (see Tab. \ref{table:lora}). 
On the other hand, higher $\text{SF}$ results in higher receiver sensitivity (often below the noise floor) thus extending the communication range from the gateway and improving outage performance in the absence of any interference.
Spreading factors are typically set by the NetServer by sending SNR link margin feedback in response to short test frames sent out by end-devices after it successfully joins a network \cite{sornin2015lorawan}.
The corresponding ranges for each $\text{SF}$ are symbolically represented in the last column on Tab. \ref{table:lora} and depend on the specific wireless propagation model, the environment, etc.

\begin{table}[t]
\renewcommand{\arraystretch}{1.3}
\caption{LoRa Characteristics of a 25 byte Message at $\text{\text{BW}}\!=\!125$ kHz }
\label{table:lora}
\centering
\scriptsize
\begin{tabular}{| c | c | c | c | c | c | c |}
\hline
\bfseries  & \bfseries bit-rate & \bfseries Packet air- & \bfseries Transmits & \bfseries Receiver & \bfseries SNR $q_{\text{SF}}$& \bfseries Range \\
\bfseries $\text{SF}$ & \bfseries kb/s & \bfseries time ms & \bfseries per hour & \bfseries Sensitivity & \bfseries dBm & \bfseries km \\
\hline\hline
7 & 5.47 & 36.6 & 98 & -123 dBm & -6 & $l_0\!-\!l_1$ \\ \hline
8 & 3.13 & 64   & 56 & -126 & -9 & $l_1\! - \!l_2$ \\ \hline
9 & 1.76 & 113  & 31 & -129 & -12 & $l_2 \!- \!l_3$ \\ \hline
10 & 0.98 & 204 & 17 & -132 & -15 & $l_3 \!-\! l_4$ \\ \hline
11 & 0.54 & 372 & 9  & -134.5 & -17.5 & $l_4 \!-\! l_5$ \\ \hline
12 & 0.29 & 682 & 5  & -137 &  -20 & $>\!l_5$ \\ \hline

\end{tabular}
\end{table}
\normalsize

The LoRa MAC layer is basically an ALOHA protocol controlled by the NetServer. 
Significantly, the gateways can receive signals from multiple (currently up to 9 \cite{vangelista2015long}) end-devices simultaneously due to the orthogonality of transmission sub-bands and {quasi-orthogonality of} SFs.
If for example two or more signals are received at the same time, at the same frequency, and same SF, the FFT output at the gateway would produce two or more indiscernible peaks.
To this end, it is estimated that gateways can successfully receive colliding packets with equal SFs if the desired signal is at least 6 dB stronger than any other.
On the other hand, collisions of signals of different SFs are practically orthogonal since the rejection gain ranges from 16 to 36 dB \cite{goursaud2015dedicated}. 
We will therefore not consider inter-spreading factor interference  and instead focus on co-spreading factor interference as described below.

\squeezeup
\section{Single Gateway: Uplink System Model}

\begin{figure}[t]
\centering
\includegraphics[scale=0.25]{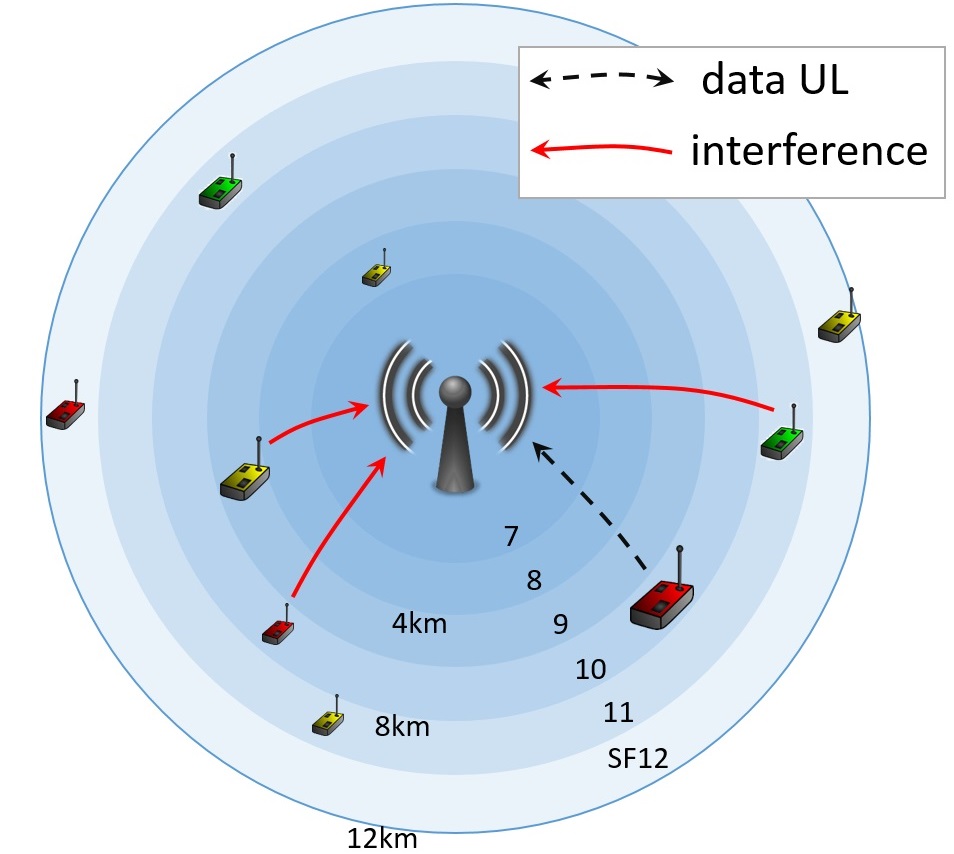}
\caption{
System setup in the uplink consisting of just one gateway, and several concurrently transmitting end-devices located uniformly in a radius of $R$ km.}
\label{fig0}
\end{figure}

In this section, we model the uplink to a single gateway taking into account possible interference from colliding signals.
Particular emphasis is given to the spatial distribution of the end-devices by leveraging tools from Stochastic Geometry \cite{haenggi2012stochastic}.
Namely, we consider a gateway located at the origin of the coordinate system and end-devices uniformly located at random in some deployment region $\V\!\subseteq\!\R^2$ described through an inhomogeneous Poisson point process (PPP) $\Phi$, with intensity function $\rho\!>\! 0$ in $\V$, and $0$ otherwise.
Each point of the PPP represents an end-device.
For simplicity we assume that $\V$ is a disk of radius $R$km of area $V\!=\!|\V|\!=\!\pi R^2$ containing a total of $N$ end-devices, where $N$ is a Poisson distributed random variable with mean $\bar{N}\!=\!\rho V$.
The Euclidean distance from end-device $i$ to the gateway at the origin is denoted as $d_i$km.
Devices transmit in the uplink (UL) at random using ALOHA and also satisfy an additional maximum $p_0\!=\! 1\%$ duty cycling policy as specified by ETSI \cite{sornin2015lorawan}.
Therefore, end-devices using higher SFs will transmit less often as to respect this policy.
We assume that all transmissions occur in a single BW=125kHz channel for simplicity.
Regardless of this simplification, {concurrently received signals of the same frequency and SF interfere at the gateway} and may result in severe packet losses which need to be catered for by suitable retransmission mechanisms, thus wasting away valuable battery power, while also incurring end-to-end delays and additional signalling overheads.
Finally, we assume that SFs are assigned by the NetServer according to the distances $d_i$ from the gateway as described in Tab. \ref{table:lora}.
A schematic of the setup is shown in Fig.\ref{fig0}.

We focus on a single end-device and investigate the co-spreading factor interference caused by simultaneous transmissions.
Let the desired signal from an end-device denoted by $s_1(t)$ be transmitted over a block flat-fading channel $h(t)$ (quasi-static) modelled as a zero-mean, independent, circularly-symmetric complex Gaussian random variable with unit variance (i.e., Rayleigh fading). 
The received signal at the gateway can therefore be described by
$
r_1(t) \!=\! g(d_1) h_1(t) \!\ast\! s_1(t)\! +\! \sum_{k=2}^N \! { \chi_k^{\text{SF}}(t) } g(d_k) h_k(t) \!\ast\! s_k(t) \!+\!  n(t)
$, 
where $n(t)$ is additive white Gaussian noise (AWGN) with zero-mean and variance $\mathcal{N}\!=\! -174\!+\! \text{NF} \!+\! 10\log\text{BW}$ dBm, NF is the receiver noise figure and is fixed for a given hardware implementation, here taken to be 6 dB, {$\chi_k^{\text{SF}}(t)$ is the indicator function, indicating if a different end-device, denoted by a subscript $k\!>\!1$, is transmitting at the same time, frequency and SF as the desired end-device thus causing interference}, and $g(d_i)$ is the path loss attenuation function defined as
$
g(d_i) \!= \! (\lambda / 4 \pi d_i)^\eta ,
$
which follows from the Friis transmission equation, where $\lambda$ is the carrier wavelength, and $\eta\!\geq\! 2$ is the path loss exponent usually taken to be equal to (2.7) 4 in (sub-) urban environments.
For the sake of simplicity we have assumed isotropically radiating antennas at both transmitter and receiver ends {and have assumed negligible interfering signals from non-LoRa signals operating in ISM frequencies.}

\squeezeup
\section{Uplink Outage Probability}

Outage of a desired signal in the uplink can occur at the gateway if either of these conditions are satisfied: 
\begin{enumerate}
\item{the received signal to noise ratio (SNR) is below the $\text{SF}$ specific threshold $q_{\text{SF}}$ (see Tab.\ref{table:lora});}
\item{the received signal is less than four times (6 dB) stronger than any other concurrent transmission of the same $\text{SF}$.}
\end{enumerate}

\subsubsection{Outage Condition 1}
The instantaneous SNR can be defined as
$
\text{SNR} = \mathcal{P}_1 |h_1|^2 g(d_1) /\mathcal{N}
$,
where $\mathcal{P}_i $ is the transmit power of end-device $i$ in milliwatts, and $|h_i|^2$ is the channel gain modelled as an exponential random variable with mean one.
We can thus formulate the first outage condition as the complement of the connection probability:
\es{
H_1= \mathbb{P}\big[\text{SNR} \geq q_{\text{SF}} \big| \, d_1 \big],
\label{out1}}
which essentially calculates the probability that at any given instance of time, a received signal $s_1(t)$ from an end-device located $d_1$km from the gateway will not satisfy the SNR threshold $q_\text{SF}$, a piecewise constant function of the distance $d_1$ as described in the penultimate column of Tab. \ref{table:lora}.

\subsubsection{Outage Condition 2}
The second outage condition is concerned with the strongest interfering received signal which is of the same spreading factor as the desired one.
We therefore label the strongest interfering signal $k^*$ defined as
\es{
k^*= \arg\max_{k>1}\{ \mathcal{P}_k \chi_k^{\text{SF}} |h_k|^2 g(d_k) \}
,\label{ou}}
where we have dropped the time dependence of received signals since the system is assumed ergodic (i.e., any two instances of time are statistically independent).
Note that the transmit powers of end-devices with the same SF signals are assumed equal.
The second outage condition is therefore given by the complement of:
\es{
Q_1 = \mathbb{P} \Big[ \frac{ |h_1|^2 g(d_1)}{ |h_{k^*}\!|^2 g(d_{k^*}\!)}   \geq 4  \, \Big| \, d_1 \Big]
\label{out2},}
{thus providing a statistically meaningful performance metric quantifying when collisions of the same SF are significant.
Intuitively, we expect $Q_1$ to decay with increasing $\bar{N}$.}

Combined, the two outage conditions form the joint outage probability $J_1$ of a received signal $s_1$ given by the complement of a successfully received signal {defined as $J_1\!=\! 1\!-\! H_1 Q_1$}.

\subsubsection{Coverage Probability}
{The coverage probability is the probability that a randomly selected end-device is in coverage (i.e., not in outage) at any particular instance of time.}
One may obtain the system's coverage probability $\wp_c$ with respect to $\mathcal{X}\!=\!\{H_1,Q_1,H_1 Q_1 \}$ by de-conditioning on the position of the specific end-device achieved by averaging over $\V$
{
\es{
\wp_c [\mathcal{X}] \!=\! \frac{2}{R^2} \int_0^R \! \mathcal{X} (d_1)  d_1 \dd d_1,
\label{cov}}
thus giving a system-level performance metric for a single gateway LoRa network with approximately $\bar{N}$ end-devices in terms of the complementary outage probability.
Of course, $\wp_c[H_1]$ is independent of the deployment density $\rho\!=\! \bar{N}/V$.
}

{
\squeezeup
\squeezeup
\subsection{Mathematical Analysis}
\subsubsection{Outage Condition 1}
We can directly calculate \eqref{out1} by simply rearranging SNR for $|h_1|^2 \!\sim \!\exp(1)$ to get
\es{
H_1(d_1)=  \mathbb{P}\prr{|h_1|^2 \geq \frac{ \mathcal{N} q_{\text{SF}} }{\mathcal{P}_1 g(d_1)} \, \Big| \,  d_1}
\! =\! \exp\pr{- \frac{ \mathcal{N} q_{\text{SF}} }{\mathcal{P}_1 g(d_1)}}
\label{out11}.
}
Note that other than the distance dependent outage condition $q_{\text{SF}}$, equation \eqref{out11} is the standard point-to-point complementary outage probability and can be calculated for other wireless fading channels \cite{yacoub2007alpha}, anisotropic antenna gains \cite{georgiou2013connectivity}, and for MIMO arrangements \cite{coon2015connectivity}.
Moreover, note that \eqref{out11} is independent of the end-device deployment density $\rho\!=\!\bar{N}/V$.}

\subsubsection{Outage Condition 2}
The network performance analysis due to co-spreading factor interference, as embodied by \eqref{ou} and \eqref{out2} is non-standard and novel.
To calculate the second outage condition through \eqref{out2} we make use of the theory of order statistics (maximum among several i.i.d. random variables)
\es{
Q_1(d_1) = \mathbb{E}_{|h_1|^2}\prr{ \mathbb{P} 
\Big[ X_{k^*} < |h_1|^2 g(d_1)/4   \, \Big |\, |h_1|^2 , d_1 \Big] }
}
where we have set $X_{k^*}\!=\! |h_{k^*}|^2 g(d_{k^*})$.
To make progress we first require the product distribution of $X_i \!=\! |h_i|^2 g(d_i)$ which we now calculate for the case of a uniform deployment of $N$ end-devices in a disk of radius $R$ km around the gateway.

\textit{Product distribution:}
We assume that only end-devices located inside an annulus $\hat{\V}(d_1)\subset \V$ defined by the inner and outer radii $l_j$ and $l_{j+1}$ km, respectively, have the same $\text{SF}$ as the desired signal from the end-device located at $d_1\in[l_j,l_{j+1})$.
We therefore have that $|\hat{\V}(d_1)|\!=\! \pi (l_{j+1}^2 -l_j^2) $.
Therefore, the pdf of the distance $d_i$ to the gateway of a randomly chosen point $i$ within the same annulus $\hat{\V}(d_1)$ is given by $f_{d_i}(x) = 2\pi x /|\hat{\V}(d_1)|$. 
Calculating the pdf of $g(d_i)$ 
\es{
f_{g(d_i)} (x) = \Big| \frac{\dd}{\dd x}g^{-1}(x) \Big| f_{d_i}\big(g^{-1}(x)\big) =\frac{\lambda^2 x^{-\frac{\eta+2}{\eta}}}{8\eta \pi |\hat{\V}(d_1)|}
} 
which has a finite support on $g(l_{j+1}) \! \leq \! x \!\leq\! g(l_j)$, and recalling that $|h_i|^2 \!\sim\! \exp(1)$, it follows that the pdf of $X_i$ is  
\es{
f_{X_i}(z) &= \int_{g(l_{j+1})}^{g(l_j)} \frac{1}{x} f_{g(d_i)}(x) f_{|h_i|^2}(z/x) \dd x \\
&=  \frac{\lambda^2 z^{-\frac{\eta+2}{\eta}}}{8 \eta \pi |\hat{\V}(d_1)|} \prr{ \Gamma \Big(1+\frac{2}{\eta } , \frac{z}{g(x)}   \Big) }_{x=l_{j+1}}^{x=l_j }  ,
\label{Xi}}
supported on $z\in \R^+$, where $\Gamma(\cdot,\cdot)$ is the upper incomplete gamma function.
Integrating \eqref{Xi} we arrive at the cdf of $X_i$ 
\es{F_{X_i}(z) \!=\! \frac{z^{\frac{2}{\eta}}\lambda^2}{16 \pi |\hat{\V}(d_1)|} \bigg[ \frac{(e^{\frac{-z}{g(x)}}\!-\!1)z^\frac{2}{\eta}}{g(x)^\frac{2}{\eta}}\!-\!\Gamma\Big(1\!+\!\frac{2}{\eta},\frac{z}{g(x)}\Big)\bigg]_{x=l_{j+1}}^{x=l_j}
\label{FX}}

\begin{figure}[t]
\centering
\includegraphics[scale=0.21]{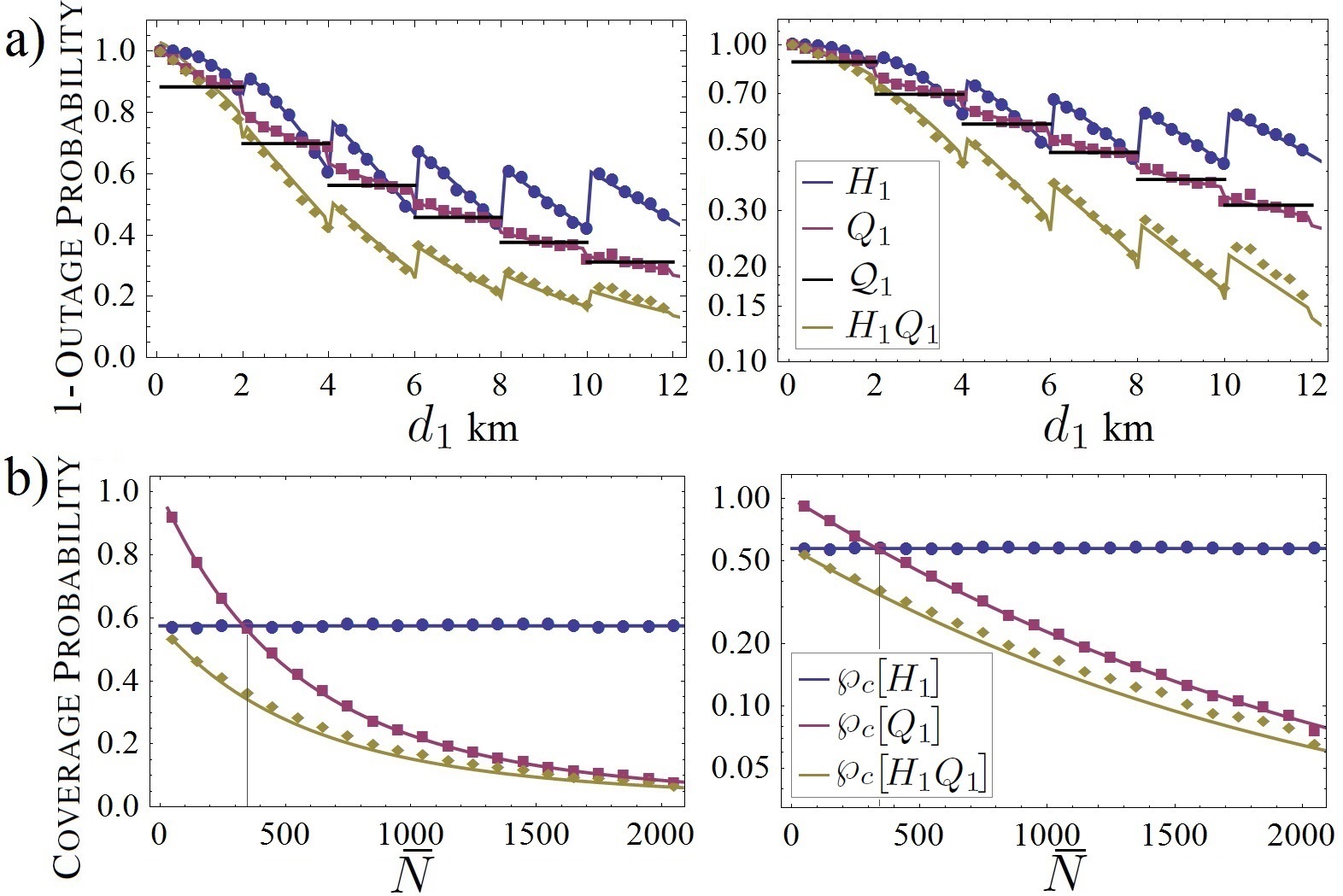}
\caption{
{a) Complement of the outage probabilities $H_1$ (blue), $Q_1$ (purple), $\mathcal{Q}_1$ (black), and $H_1\!\times\!Q_1$ (yellow) plotted as functions of the distance from the gateway $d_1$km assuming an average of $\bar{N}\!=\!500$ end-devices in a deployment area of radius $R\!=\!12$km.
b) Coverage probabilities $\wp[\mathcal{X}]$ for $\mathcal{X}\!=\!\{H_1,Q_1,H_1 Q_1\}$ using the same colouring and markers as in a) for different mean values of end-devices $\bar{N}\!\in\![1,2000]$. 
Solid lines are calculated via \eqref{out11}, \eqref{Q1}, and \eqref{Ma}, and numerically integrated according to \eqref{cov}, whilst markers are obtained via Monte Carlo simulations.
The right panels are the same as on the left but plotted on a log-linear scale. 
Parameters used: $\eta\!=\!2.7$ (sub-urban), $p_0 \!=\! 1\%$, and $\mathcal{P}_1 \!=\! 19$dBm.
}}
\label{fig1}
\end{figure}

\squeezeup
\textit{Order statistics:}
From a sample of $n>0$ independent and identically distributed random variables distributed according to $F_{X_i}(z)$, we may obtain the distribution of the maximum, i.e., the strongest interfering signal $X_{k^*}$, by using the theory of order statistics:
$
F_{X_{k^*}}(z)\!=\!\mathbb{E}_n \Big[ [F_{X_i}(z)]^n \Big]
$, 
where the sample size $n$ is a Poisson distributed random variable with mean $v\!=\!p_0 \rho |\hat{\V}(d_1)|$ given by the expected number of concurrently transmitting end-devices in the same SF annulus $\hat{\V}(d_1)$ as that of the desired signal.
Using these definitions we can write 
$
F_{X_{k^*}}(x) \!=\! \sum_{k=0}^\infty [F_{X_{i}}(x)]^{k} \frac{v^k e^{-v}}{k!}
.$
Deconditioning on the channel gain $|h_1|^2$ we finally arrive at
\es{
Q_1 \!=\! \mathbb{E}_{|h_1|^2}\Big[ F_{X_{k^*}}\Big( \frac{|h_1|^2 g(d_1)}{4}\Big) \Big] \!=\! \int_0^\infty \!\! e^{-z}  F_{X_{k^*}}\Big(\frac{z g(d_1)}{4} \Big) \dd z
\label{Q1}
.}
Equation \eqref{Q1} can only be computed numerically. 
Instead we may approximate it by Taylor expanding $F_{X_{i}}(z g(l_{j+1})/4)$ for small $z\ll 1$, and retaining the leading order term
to obtain a rough approximation of $Q_1(d_1)$ in closed form given by
\es{
\mathcal{Q}_1 \!\approx\! \frac{2e^{-v} l_{j+1}^{\eta} (\eta +2) |\hat{\V}(d_1)| }{\pi v l_j^{\eta+2} + l_{j+1}^\eta \big( 2(\eta+2)|\hat{\V}(d_1)| - \pi v\l_{j+1}^2  \big) }
\label{Ma}}
{Note that $\mathcal{Q}_1$ has a piecewise constant dependence on $d_1$ via $l_j$, $l_{j+1}$, $v$, and $|\hat{\V}(d_1)|$, and is therefore a very crude approximation of $Q_1$ as can be seen from the numerical simulations in Fig. \ref{fig1}a) described below.
Nevertheless, \eqref{Ma} captures the general trend of $Q_1$ as confirmed by numerical simulations, and is much easier to calculate than \eqref{Q1}.
Moreover, note that this general trend can often be more insightful and practically helpful for wireless network design and field engineers. 
}

\squeezeup
\squeezeup
\subsection{Numerical Simulations and Discussion \label{num}}
Fig. \ref{fig1} shows Monte Carlo computer simulation results verifying the above derivations.
For simplicity, we use Semtech's recommended values of $l_i\!=\!2i$ for $i\!=\!0,\ldots5$  \cite{LoRaWAN}.
Each marker in Fig. \ref{fig1} corresponds to the simulated performance of the single gateway LoRa network in the UL, averaged over $10^5$ random deployment realisations of the PPP in $\V$.
An excellent agreement is observed between the derived results and the simulated ones, except for $\mathcal{Q}_1$ which only captures the downward staircase trend of $Q_1$.
A distance dependent SNR threshold $q_{\text{SF}}$ is assumed (see penultimate column of Tab.\ref{table:lora}).
This has a striking saw-tooth effect on the SNR dependent outage condition $H_1$, demonstrating a boost in performance as an end-device transitions into regions of higher $\text{SF}$. 
This is a unique feature of LoRa and is a direct consequence of $q_{\text{SF}}$.
Interestingly however, the saw-tooth direction is reversed and the boost becomes a drop when considering co-spreading interference in $Q_1$. 
This behaviour is purely due to geometrical reasons. 
Namely, for uniform PPP the number of interfering end-devices in adjacent SF regions is proportional to $|\hat{\mathcal{V}}(d_1)|\!\sim \!d_1$.
Hence the saw-tooth boosting effect is somewhat diluted when considering the joint complementary outage probability $H_1 Q_1$ (yellow curve). 
Finally, it is observed that coverage probabilities $\wp_c[Q_1]$, and $\wp_c[H_1 Q_1]$ decays exponentially with the expected number of end-devices $\bar{N}$ whilst $\wp[H_1]$ is constant as expected (see right panel of Fig. \ref{fig1}b)).
This is a direct consequence of co-spreading factor interference where it becomes increasingly less likely that the desired signal is at least four times stronger than any of the interfering ones.
Interestingly, it is possible to distinguish when co-spreading factor interference is the dominant cause of outage, i.e., a scalability limit, which in Fig. \ref{fig1}b) is indicated by a vertical line.
This of course depends strongly on the wireless propagation environment and the transmission scheme details.

\squeezeup
\section{Conclusion}

We have investigated the effects of interference in a single gateway LoRa network, a LPWA technology with promising IoT applications.
Unlike other wireless networks, LoRa employs an adaptive CSS modulation scheme, thus extending the communication range in the absence of any interference.
Interference is however present when signals simultaneously collide in time, frequency, and spreading factor.
Leveraging tools from stochastic geometry, we have formulated and solved two link-outage conditions, one based on SNR, and the other on co-spreading sequence interference.
Each displays interesting behaviours, unique to LoRa, with the latter causing performance to decay exponentially with the number of end-devices, despite various interference mitigation measures available to LoRa, thus limiting its scalability.
It is interesting that LoRa networks appear to be impervious to cumulative interference effects (typically modelled as shot-noise \cite{haenggi2012stochastic}). 
If this assumption is invalid, then our qualitative results are simply optimistic upper bounds towards network scalability.
Going beyond this first foray into the modelling of LoRa, it would be interesting to understand the effets of multiple gateways \cite{augustin2016study}, and spatially inhomogeneous deployments.
Finally, we point towards recently developed packet-level simulators \cite{LoRaSim} which can further shed light into the performance of LoRa networks.


\squeezeup
\bibliographystyle{ieeetr}
\bibliography{mybib}

\end{document}